# Rigid multibody simulation of a helix-like structure –
## *The dynamics of bacterial adhesion pili*

Johan Zakrisson[a], Krister Wiklund[a], Martin Servin[a], Ove Axner[a,c], Claude Lacoursière[b], Magnus Andersson[a,c,*]

Department of [a]Physics and [b]Computer Science, [c]Umeå Centre for Microbial Research (UCMR), Umeå University, SE-90187, Umeå, Sweden.

*Corresponding author: Magnus Andersson, Department of Physics, Umeå University, SE-901 87 Umeå, Sweden, Tel: + 46 – 90 786 6336, Email: magnus.andersson@physics.umu.se

**Running title:** Macromolecule multibody simulation

## ABSTRACT

We present a coarse grained rigid multibody model of a subunit assembled helix-like polymer, e.g., adhesion pili expressed by bacteria, that is capable of describing the polymer's force-extension response. With building blocks representing individual subunits the model appropriately describes the complex behavior of pili expressed by the gram-negative uropathogenic *Escherichia coli* bacteria under the action of an external force. Numerical simulations show that the dynamics of the model, which include both the effects of unwinding and rewinding, are in good quantitative agreement with the characteristic force-extension response as observed experimentally for type 1 and P pili. By tuning the model, it is also possible to reproduce the force-extension response in the presence of anti-shaft antibodies, which dramatically changes the mechanical properties. Thus, the model and the results in this work give enhanced understanding of how a pilus unwinds under action of external forces and provide new perspective of the complex bacterial adhesion processes.

## INTRODUCTION

In an era of increasing antibiotic resistance, enhancing the understanding of bacterial adhesion is important both from fundamental perspectives and for the development of new drugs. The adhesion process, which constitutes the initial step for a bacterium to invade a host cell, is for uropathogenic *Escherichia coli* (UPEC) and enterotoxic *Escherichia coli* (ETEC) mediated via external attachment organelles, known as pili (Sauer et al 2000). UPEC bacteria serve as a good model system for general bacterial adhesion since some of their pili have been well characterized with respect to their geometrical structure and intrinsic mechanics (Bullitt and Makowski 1998; Hahn et al 2002; Forero et al 2006; Miller et al 2006; Andersson et al 2007; Lugmaier et al 2008). Depending on the environmental cues along its journey *in vivo*, e.g. in the urethra, the bladder, or the kidney, analogous but different types of pili are expressed (Schwan et al 2002). For example, P and type 1 pili are predominantly expressed in the upper and lower urinary tract, respectively. The reason for this is assumed to be that the various types of pili have dissimilar biomechanical properties, presumably dissimilar compliances. These types of pili, which are ~7 nm thick and ~1-2 μm long, have been well described and it has been found that they possess a large amount of elasticity, mainly originating from their helix-like structure, which is assembled from a large number (~$10^3$) of subunits (Fällman et al 2004; Andersson et al 2007).

Helix-like adhesion pili such as P and type 1 pili that are exposed to an external force can unwind their quaternary structure. Consecutive layers are held together by layer-to-layer (*ll*) bonds. By breakage of these, the coil-like structure can be unwounded, whereby the pili can be elongated several times its original length. The unwinding can be divided into three distinct force regions; I) a linearly increasing force with extension, II) a constant force, and III) a non-linear force-*vs.*-elongation response, whose first part constitutes a pseudo-linear response. It is assumed that the latter response plays a minor role; the unique properties of these structures comprise the extension under a constant force in region II. Moreover, an elongation under adiabatic conditions (i.e. "slowly"), which takes place under a given lowest possible unwinding force, gives a contraction pattern during rewinding that is almost identical to the unwinding. It is therefore hypothesized that individual pili can unwind and recoil to distribute the shear force acting on the bacterium caused by the rinsing action of the urine to several pili and that this process would reduce the load on individual adhesins that are placed at the tip of the rods, and thereby prolonging the lifetime for attachment (Duncan et al 2005; Forero et al 2006; Zakrisson et al 2012; Rangel et al 2013). As shown in an experimental assay by Forero et al. (Forero et al 2006) and later by computer simulations (Zakrisson et al 2013) the force needed for unwinding is well suited for keeping an optimal load on the FimH adhesin. All this point at that the biomechanical properties of the pili help the bacterium to maintain firm adhesion to the host cells, even in the presence of considerable flow, increasing the chances for infection. This leads to the assumption that, by compromising the unwinding and rewinding capabilities of pili with antibodies, peptides or other chemical substances, pathogenic bacteria can possibility be prevented from attaching to host cells.

The fact that the biomechanical properties of adhesion pili can be compromised by an external substance was recently demonstrated by optical tweezers force spectroscopy measurements. It was found that the biomechanical properties of P pili was changed in the presence of the chaperon protein PapD (Klinth et al 2012). In addition, a study using anti-P pili antibodies showed an even more pronounced change of the mechanical properties of P pili, with a significantly reduced elasticity, where the force needed to unwind the structure increased and the length of region that extend under a constant force decreased (Mortezaei et al 2013). These results open up for an intriguing possibility that antibodies could directly interfere with the adhesion process when attaching to the shaft of pili and disable their elasticity.

A pilus is thus a feature-rich macro-molecule and a rigorous investigation of its behavior is of importance when it comes to understanding bacterial adhesion. Experimental measurements *in situ* and computer simulations, based upon different methods, can provide enhanced understanding at a micro-molecular level, i.e., of the interactions and the kinetics of bonds. We and other groups have previously performed high resolution force measurements, using optical tweezers and atomic force microscopy, which revealed macroscopic properties such as force-extension behavior and microscopic entities, e.g., transition energies and bond lengths, for few types of pili (Andersson et al 2006a; Forero et al 2006; Andersson et al 2007; Lugmaier et al 2008; Castelain et al 2011).

To elucidate the biomechanical properties of helix-like organelles, in particular when exposed to stress, a few models describing the force-extension response of helix-like pili have been developed. The first one, which was a fully analytical model, was based upon the sticky-chain concept (Andersson et al 2006b). The model reproduced the steady-state unwinding behavior of pili well, whereas it could not perfectly reproduce all aspects of the rewinding process, since the model only partly considers the complex 3D structure. An alternative model is based upon a kinetic three state Monte Carlo model (Björnham et al 2008). This model is also capable of accurately reproducing the unwinding although it can model the rewinding only to a certain extent. Hence, although models have been developed that can describe many of the complex features of helix-like pili fairly well, none is capable of describing all phenomena.

To remedy this problem, we present here a model built upon rigid multibody simulations that can elucidate the behavior of a subunit-assembled helix-like structure under strain. By its parameterization, the model incorporates geometrical shapes assessed from structural data and optical tweezers force extension experiments performed *in situ*. The model is taking into account the complex 3D structure of the pili and with well-chosen model parameters it is capable of reproducing the force-extension response of a single pilus well. It thus provides an improved understanding of the mechanics of pili and supports previous assumptions that pili are bacterial biopolymers that are capable of reducing the external load on the adhesins, which is considered important for the initial attachment and sustained adhesion. Finally, we show that this model can reproduce the force responses of pili exposed to shaft specific antibodies.

## THEORY AND MODEL

## The aim of the work

We have developed a parameterized multibody polymer model to properly describe the biomechanical responses of helix-like attachment organelles that can reproduce the behavior of the organelles when exposed to force. In particular, it has been based upon the findings that the main rigid helical rod of P and type 1 pili are composed by hundred to thousand copies of the PapA or FimA subunits (Hahn et al 2002; Mu and Bullitt 2006), that the PapA and FimA subunits are ~4.1 nm and ~5.7 nm long with a molecular mass of ~16.5 and ~15.8 kDa, respectively (Bullitt and Makowski 1998; Forero et al 2006), that the subunits are attached by head-to-tail (*ht*) bonds (strong hydrophobic interactions) and coiled into a higher order helix-like structure with ~3 subunits per turn by *ll*-bonds (Bullitt and Makowski 1998; Hahn et al 2002), and that the helix-like structure is the natural configuration of a pilus at rest. It has also been found that a short tip fibrillum that holds the adhesin that binds to receptors on host cells is attached at the end of the rod. A low-resolution reconstruction map of a P-pilus (using the structure of the PapA subunit, PDB:2UY6, (Verger et al 2007)) is shown in Fig. 1A, which illustrates both the 3D-geometrical structure (a helical rod) and a short unwound part (a linearized rod).

The model should in particular be able to reproduce results from optical tweezers force measurements of pili. A descriptive illustration of such an experiment is shown in Fig. 2A and a detailed explanation is found in Anderson et al. (Andersson et al 2008). An experiment is performed by attaching a bacterium to a large bead that is immobilized to a cover slip. Using optical tweezers, a bead is trapped and attached to a pilus. While the tweezers stay fixed, holding the bead in its focal region, the bead-distance is increased by moving the bacterium using a piezo-stage with nm resolution. The pilus is thereby exposed to an enforced elongation and by monitoring the position of the bead in the trap the resulting force-*vs.*-elongation response is measured. An example of a force-extension curve of a pilus, where the black and red curve represent unwinding and rewinding of the pilus, respectively, is shown in Fig. 2B. Note that in this figure the force is normalized with respect to the constant force plateau in region II for easy comparison with the simulation results.

## A coarse grained rigid multibody polymer model

Since the helical rod is the major part of a pilus and therefore believed to be mainly responsible for the complex and feature rich biomechanical properties of the pili, we have focused on modeling only the rod in this work. As is described below, we model an individual pilus as a helix-like structure composed of kinematically constrained homogeneous subunits whose movements are described by the Newton-Euler equations of motion. Based on data from the literature, the free parameters in the model are adjusted so its predictions agree with experimental findings.

As is shown in Fig. 1B, the helical rod is modeled as a discrete system of $N$ rigid bodies, each representing a subunit connected in a chain topology by $N$-$1$ $ht$-bonds and $N$-$3$ $ll$-bonds. The one-dimensional position and orientation (quaternion or Euler angles) of each subunit $i = 1,...,N$ are denoted by $x_i$ and $q_i$, respectively. The total potential energy for the model is assumed to be given by a sum of the potential energies of all the $ht$-bonds, $U_{i,i+1}^{ht}$, all the $ll$-bonds, $U_{i,i+3}^{ll}$, and that of the bond between the last sub-unit and the measurement bead, $U_N^{bead}$, i.e. as,

$$U = \sum_{i=1}^{N-1} U_{i,i+1}^{ht} + \sum_{i=1}^{N-3} U_{i,i+3}^{ll} + U_N^{bead}, \tag{1}$$

The $ht$-bonds resist stretching and rotation, both bending and twisting, in the joint connecting each subunit pair of rigid bodies. We model this bond as a sum of a stretch and rotation, i.e. as

$$U_{i,i+1}^{ht} = U_{i,i+1}^{stretch} + U_{i,i+1}^{rotation} \tag{2}$$

The potential of the former is assumed to have a quadratic dependence on elongation, giving rise to a force that is linear with elongation. Hence, it is modelled as

$$U_{i,i+1}^{stretch} \equiv \tfrac{1}{2} \kappa_{stretch} \delta_{i,i+1}^2, \tag{3}$$

where $\delta_{i,i+1} = \delta(r_i, q_i, r_{i+1}, q_{i+1})$ denotes the separation distance between the points in body $i$ and $i+1$ at the center for that $ht$-bond. In order to provide a restoring force around the equilibrium position the potential energy due to rotation is described as a polynomial of even powers of rotation angels, *viz.* as,

$$U_{i,i+1}^{rotation} \equiv \sum_{n=1}^{k} \tfrac{1}{2n} \kappa_{\theta,2n} \theta_{i,i+1}^{2n} + \tfrac{1}{2n} \kappa_{\varphi,2n} \varphi_{i,i+1}^{2n} + \tfrac{1}{2n} \kappa_{\psi,2n} \psi_{i,i+1}^{2n}, \tag{4}$$

where $n$ is an integer and $2k$ is the order of the polynomial. The angles $\theta_{i,i+1}$, $\varphi_{i,i+1}$, and $\psi_{i,i+1}$ are the relative orientation between pairs of subunits in relation to a rest-state orientation, where $\theta_{i,i+1}$ is the angle between the local axes of the two subunits (bending), $\varphi_{i,i+1}$ is the angle of rotation around the axis of the subunit i, and $\psi_{i,i+1}$ is the angle of rotation around the axis of subunit $i+1$. In each subunit, the orientation angles are computed given three mutually orthogonal body-fixed direction vectors. The various $\kappa$-parameters are potential stiffness parameters, which are chosen so the model matches experimental data. In the future, the rotation potential can be replaced with any kind of functional dependence on orientations whenever such have been determined from experiments.

The physical description of the interactions in the $ll$-bond is complex and not known in detail. We model the interaction by assigning point "charges" to the subunits and an interaction potential that we give a characteristic length scale, represented by the distance to the transition barrier, $L_0$, and a potential depth, $U_0^{ll}$, which may be determined from experiments. More precisely, as is illustrated Fig. 1B, two charges are placed on each subunit $i$, approximately centered on two of the sides, in such a way that the positions of each coincide with one of the charges on each of the $i$-$3$ and $i$+$3$ subunits in the helical rest configuration. Due to screening effects not all charges interact, thus we only account for the $i-3$ to $i$ and the $i$ to $i+3$ interactions. The interaction potential is chosen as,

$$U_{i,i+3}^{ll} = \tfrac{1}{2} U_0^{ll} \exp\left[ -\mid r_i - r_{i+3} \mid^2 / L_0^2 \right], \tag{5}$$

where $r_i$ and $r_{i+3}$ are the positions of the interacting charges on the units $i$ and $i+3$, respectively, as is shown in Fig. 1B.

The potential $U_N^{bead}$, finally, models the coupling between the end-point of subunit $N$ and the bead, i.e., the optical trap, which has a given time-dependent position. This model uses the same building blocks as for the $ht$-bond, but with dissimilar values of the stiffness parameter; the stretch potential is assumed to have a stiffness $\kappa_{trap}$.

Moreover, the rotational potential of the last subunit is assumed to be infinitely weak. This implies that the subunit can rotate freely with respect to the trap. The coupling to the fixed bacterium is modeled by constraining the subunit 1 to have a fixed position and orientation relative to the world frame.

To achieve realistic conditions and to simulate the fluid-rod interaction we added viscous friction to the system of Newton-Euler equations by the use of viscous friction forces and torques that are linear in velocity and angular velocity, respectively.

**Large time steps by regularized constraints**

Since the system is weakly damped, individual units in the model will perform movements on both slow and fast time scales; the former are dictated by the slow movements of the coarse structure of the organelle while the latter comprise rapid oscillations around the equilibrium positions. This makes simulations based on conventional solving methodologies very time consuming. To reduce computational time and improve numerical stability, we advance the simulation in discrete time-steps modeling the dynamics of the coarse structure of the system by treating the interaction potentials as regularized constraints (Lacoursière 2007). This approach filters out dynamics on the time scale of subunit oscillation modes, while still capturing correctly the dynamics of unwinding and rewinding of the pilus. This regularized constraint makes it possible to have arbitrarily stiff potentials while keeping large time steps, without introducing significant amount of artificial damping on the slow modes. The regularization procedure for multibody systems is thoroughly described by Lacoursière (Lacoursière 2007) and have previously been applied to a system similar to the one considered here (Servin and Lacoursière 2007). In short, incorporating the potentials as regularized constraints transforms the equations of motion from a set of ordinary differential equations to a set of differential algebraic equations. The system variables are then extended to include also the Lagrange multipliers associated with the constraints. Observe that we replace all potentials, except the softer layer-to-layer potentials, by regularized constraints. After regularization and symplectic Euler discretization, the problem of integrating the equations of motions one time-step is reduced essentially to solving a sparse and narrow banded linear system of equations of size $6(N-1) \times 6(N-1)$. This implies that the computation scales linearly with the number of subunits.

## Simulation procedure

The model described above was implemented and complied as a MEX file in MATLAB. The computational time per time-step and per subunit on a computer (Intel® Core™ i7 3770 / 3.40GHz, 8GB DDR3 1600MHz RAM) was 50 µs. In a typical extension simulation of pili, the pilus was initially in its equilibrium configuration, i.e., coiled in a helix-like structure and attached with the left side to an immobilized wall representing the bacterial cell. The last subunit was attached to a bead whose position and movement are controlled by the program. The bead was moved at a constant velocity until the pilus was stretched to 85% of the theoretical full length. The bead was then moved in the opposite direction to simulate the rewinding process. At each time-step numerical values of the pili extension and the trap force were acquired. Numerical force-extension simulations were performed for a large number of pili configurations, while the number of subunits and the strengths of the interaction potentials were varied, until good agreement with measured data was achieved.

# RESULTS

## Force-extension of the rigid multibody model

We used the rigid multibody model and numerical simulation method presented above to study the behavior of a helix-like polymer under extension analogous to optical tweezers experiments. A tensile force was thereby applied at the end of the rigid body model simulating an optical trap with slightly exaggerated trap stiffness (corresponding to ~1000 pN/µm). Although the trap stiffness for optical tweezers in general is lower than this, the model parameter was given this value to avoid oscillations in the computer simulation. It was found that this did not affect the result of the simulations. To ensure proper behavior during a simulation a series of snap-shots were captured to qualitatively inspect the behavior of the subunits and the overall behavior of the structure under tension.

Subunits were model as $5 \times 1 \times 1$ nm blocks that were arranged into layers with 3.36 subunits per turn. Since the pilus' subunits are attached in a head-to-tail configuration with strong hydrophobic interactions with limited flexibility the $\kappa_{stretch}$ parameter in Eq. 3 was set to infinity, thus preventing any elastic stretching between subunits. To find appropriate parameter values for the model, i.e. the $\kappa_{\theta,2n}$, $\kappa_{\varphi+2n}$, and $\kappa_{\psi+2n}$ in Eq. 4 and $U_0^{ll}$ and $L_0$ in Eq. 5, a number of simulations were performed for a variety of parameter values until adequate agreement with the experimental data was obtained.

As is shown below, the simulation represents the experimental force-extension data well for a given set of $\kappa_{\theta,2n}$, $\kappa_{\varphi+2n}$, and $\kappa_{\psi+2n}$ value, *viz.* those that are given in Table 1, and for an *ll*-potential with an $L_0$ value of 3.5 nm. A plot of each rotation potential, i.e., in the $\theta$, $\varphi$, and $\psi$ directions, respectively, is shown in Fig. 3, describing the energy landscape for a single *ht*-bond. The inset shows a bulge in the energy landscape at ~0.6 rad, which provides the force needed to rewind the structure. As discussed below, to simplify the simulations, we exaggerated the mass of the subunits to reduce high frequency oscillations in the system. This implies that the energy potential $U_0^{ll}$ and the rotation potential $\kappa$ also are exaggerated. This is not a problem since this only affects the scaling of the force but not the dynamics of the whole system. Figure 4A shows a typical data curve from a simulation with a 250 subunit long pilus and structural parameter values similar of type 1 pili. The force in the figure is normalized with respect to the constant force plateau in region II, which has been found to be 30 pN (Andersson et al 2007).

Illustrations of different configurations of the pilus for selected parts of the elongations-retraction cycle illustrated in Fig. 4A, represented by snap-shots of the simulation, are schematically shown in Fig. 4B and further described by Table 2. For reasons of clarity, the snap-shots displayed in Fig. 4B represent solely 30 subunits. This will not markedly affect the simulation though, since the characteristic force-extension response of a pilus is independent of the number of subunits. The number of subunits only affects the slope of the region I and III, and the length of region II. A video sequence of a simulation is found in the supplementary movie S1.

The results of the simulations are in general in good agreement with the experimental data presented in Fig. 2B. The model can thus reproduce the characteristic force-extension response of the three regions seen in an experiment as well as the occurrence of the dip seen in the contraction phase.

In agreement with experimental findings (Fällman et al 2005) the simulations show a certain amount of hysteresis in region II, i.e., the unwinding force is higher than the rewinding force plateau. This is partially attributed to viscous friction introduced in the model. Since this friction is velocity dependent the hysteresis increases with extension and contraction velocity. The hysteresis is also strongly dependent on the choice of model parameters, e.g., short *ll*-bonds and weak rotational potential of the *ht*-bonds at small angles between subunits relative the rest orientation give large differences between the unwinding and rewinding force. This is supported by the observation that the hysteresis effect decreases with increased subunit masses. Since the pili extension/contraction is in general a slow, almost adiabatic process, the system response is not critically depending on the inertia of a subunit.

In addition, the simulations demonstrate a small but finite slope of region II. The slope increases with the number of subunits, but decreases with decreasing mass of the subunit elements and time step size. The former two behaviours are due to finite inertia of the system while the latter one (the dependence on the time step) originates from the numerical method.

## Reproducing force-extension data in the presence of antibodies

In order to assess to which extent the model can reproduce force-extension experiments also in the presence of antibodies, and possibly support the hypothesis that antibodies lock various layers to each other and block *ll*-interactions (Mortezaei et al 2013), we tuned the model by changing the bond potentials but kept the rest of the parameters constant. However, since the experimental data were assessed with anti-P pili antibodies and P pili we had to change the parameters of the rigid body model to represent P pili. Since P pili unwinds at 28 pN and the number of layers per layer is 3.28, while type 1 unwinds at 30 pN and has 3.36 subunits per layer, the model parameters needed only to be changed slightly when P pili were modelled. The resulting parameter values are given in Table 3.

A typical experimental force-extension response of a P pilus in the presence of antibodies is seen in Figure 5A. A comparison with data from experiments in the absence of antibodies, e.g., that given by Fig. 2B, shows that attachment of antibodies significant alters the force-extension response, mainly the constant force plateau is replaced by several force peaks indicating that the intrinsic elasticity is inhibited thus removing the important function of keeping a constant load on the adhesin.

Figure 5B shows a simulation of the rigid multibody model producing a response similar to that of Fig. 5A, i.e. for a P pilus exposed to antibodies that can lock various layers to each other and block *ll*-bonds. A snapshot of the rigid multibody model where the simulated antibodies have been attached onto the pili is presented in Fig. 5C. Again, the force is normalized with respect to the constant force plateau in region II (28 pN) (Andersson et al 2006c). To represent attachment of antibodies to the shaft subunits, a random number of *ll*-bonds were selected in the simulations and their potential strength was, during the unwinding, randomly increased 1.5-2.5 times (as compared to the normal strength). This resembles the clamping of a layer with an antibody requiring a larger force to break a *ll*-bond. As is shown by the red solid line in Fig. 5A, the force curve indicates that, during rewinding of a P pilus in the presence of antibodies, the structure is not capable of regaining its helical form. The normalized contraction force is in general <0.5 and reaches occasionally down to 0.3. We found that by decreasing the *ll*-bond strength five times the simulation could well represent the experimental data.

## DISCUSSION AND CONCLUSION

The aim of the present work was to develop a coarse grained 3D rigid multibody model that mimics the force response of subunit assembled biopolymers under tensile force. Using structural data we modeled bacterial adhesion pili commonly found on UPEC bacteria, type 1 and P pili, which are composed of numerous subunits that are ordered into a helix-like structure. Pili subunits were modeled as rigid bodies with mass, interconnected by *ht*-bonds, and twisted into a helix-like structure held together by *ll*-bonds. The motion of the bodies were modeled using the Newton-Euler equations and the parameter values were optimized until the model could reproduce a god fit to experimental data assessed using optical tweezers. For example, the best fit was found by setting the distance in the layer potential $L_0$ to 3.5 nm, which is in agreement with what is expected by comparing to force spectroscopy data. Thus, it has been found that the model can reproduce pili force-extension curves and it is also in good agreement with the two other bio-mechanical models for pili that have been developed, which are based upon the sticky-chain and a three-state Monte Carlo concept, respectively (Andersson et al 2006b; Björnham et al 2008).

The results in this work provide further support to the previously given explanation for pili mechanics, in particular its behaviour under strain and its role in bacterial adhesion (Axner et al 2010; Rangel et al 2013). First, the three defined regions of extension can be explained and visualized with the model presented in this work. Region I is explained by elastic stretching of the bonds between layers giving rise to a Hookean force response. With increased force, the *ll*-bond of the outermost subunit will break and the last subunit will change its direction from a perpendicular to a more parallel direction relative the central axis of the structure. Since it is always easier to break the *ll*-bond of the outermost subunit than for a subunit inside the helical structure this give rise to a sequential opening of subunits. This is represented by the constant forces plateau denoted region II. Region III, which is govern by stretching of *ht*-bonds, is well reproducing what is seen in force-extension experiments. In particular, the results in this work strengthen the hypothesis that the refolding process involves first a large amount of slack of the pilus before the recoiling starts, prompted by the need of the formation of a nucleation kernel for recoiling (Lugmaier et al 2008).

A limitation of the simulation model is that there are no stochastic fluctuations. Moreover, high-frequency behaviour is filtered out and damped by the numerical integrator. The absence of these fluctuations might explain some of the extra hysteresis that we observe in the simulation. In addition, to obtain realistic computational times, the mass of the subunits had to be exaggerated to damp out high frequency oscillations. This is however not a limitation since it does not affect the overall response of the system.

In order to better comprehend the molecular mechanisms and interactions governing antibodies attached to fimbriae we deliberately altered the potentials of the *ll*-bonds, $U_0^{ll}$, in the model to assess their role for the total bio-mechanical properties of the pili. It was found that attachment of antibodies could be simulated by strengthening the interaction potentials of certain bonds, randomly chosen, by a factor of ~2, representing clamping of layers. It was moreover found that blocking of rewinding could be modelled by a weakening of the same interactions by a factor of ~5. As can be seen comparing Fig. 5A and B, simulations using these settings reproduced well the results from experiments. The possibility to reproduce force-extension experiments with this physical model, in the presence of proteins or other chemical substances that affects *ll*-bonds, opens up for interesting studies of means to reduce the ability of pili to withstand large rinsing flows in the future. For example, by also implementing a model for adhesins that depends on the applied force, it can be of use to researchers to investigate how and where to target the shaft with anti-bacterial compounds. This will be significant in the aim for developing novel drugs that remove pathogenic bacteria.

The model and the simulation method presented in this paper can also be used for further simulation studies of both pili and other similar polymer structures. For example, simulated force-extension experiments of the T4 pili expressed by *Streptococcus pneumoniae* (Castelain et al 2009), which is an adhesion organelle with subunits covalently linked but lacking the helical structure, and the significantly different (weak *ll*-bonds and high dynamics) helix-like CFA/I pili expressed by ETEC strains (Andersson et al 2012), can provide important information of how bacterial adhesion organelles of different structure function. Furthermore, the model presented is coarse grained and, despite the presence of stiff forces, the simulation method allows large time-steps (about 10 ms in these simulations) with maintained numerical stability. This approach is called for when fast simulations of the forces and dynamics on the larger spatial and temporal scales of the polymer is of main importance, and this does not depend critically on dynamics on shorter spatial and temporal scales. Therefore, we believe that the model and method presented in this work can be useful in simulation studies of bio-mechanical macromolecules.

## ACKNOWLEDGEMENTS

This work was performed within the Umeå Centre for Microbial Research (UCMR) Linnaeus Program supported from Umeå University and the Swedish Research Council (349-2007-8673) and supported by the Swedish Research Council (621-2013-5379) to M.A.

# REFERENCES

Andersson M, Axner O, Almqvist F, et al (2008) Physical properties of biopolymers assessed by optical tweezers: analysis of folding and refolding of bacterial pili. ChemPhysChem 9:221–35. doi: 10.1002/cphc.200700389

Andersson M, Björnham O, Svantesson M, et al (2012) A structural basis for sustained bacterial adhesion: biomechanical properties of CFA/I pili. J Mol Biol 415:918–28. doi: 10.1016/j.jmb.2011.12.006

Andersson M, Fällman E, Uhlin BE, Axner O (2006a) Dynamic force spectroscopy of E. coli P pili. Biophys J 91:2717–25. doi: 10.1529/biophysj.106.087429

Andersson M, Fällman E, Uhlin BE, Axner O (2006b) A sticky chain model of the elongation and unfolding of Escherichia coli P pili under stress. Biophys J 90:1521–34. doi: 10.1529/biophysj.105.074674

Andersson M, Fällman E, Uhlin BE, Axner O (2006c) Force measuring optical tweezers system for long time measurements of P pili stability. Proc SPIE 6088:286–295. doi: 10.1117/12.642206

Andersson M, Uhlin BE, Fällman E (2007) The biomechanical properties of E. coli pili for urinary tract attachment reflect the host environment. Biophys J 93:3008–14. doi: 10.1529/biophysj.107.110643

Axner O, Björnham O, Castelain M, et al (2010) Unraveling the Secrets of Bacterial Adhesion Organelles using Single Molecule Force Spectroscopy. In: Gräslund A, Rigler R, Widengren J (eds) Springer Ser. Chem. Phys. single Mol. Spectrosc. Chem. Phys. Biol. Springer Berlin Heidelberg, Berlin, Heidelberg, pp 337–362

Björnham O, Axner O, Andersson M (2008) Modeling of the elongation and retraction of Escherichia coli P pili under strain by Monte Carlo simulations. Eur Biophys J 37:381–91. doi: 10.1007/s00249-007-0223-6

Bullitt E, Makowski L (1998) Bacterial adhesion pili are heterologous assemblies of similar subunits. Biophys J 74:623–32. doi: 10.1016/S0006-3495(98)77821-X

Castelain M, Ehlers S, Klinth JE, et al (2011) Fast uncoiling kinetics of F1C pili expressed by uropathogenic Escherichia coli are revealed on a single pilus level using force-measuring optical tweezers. Eur Biophys J 40:305–16. doi: 10.1007/s00249-010-0648-1

Castelain M, Koutris E, Andersson M, et al (2009) Characterization of the biomechanical properties of T4 pili expressed by Streptococcus pneumoniae--a comparison between helix-like and open coil-like pili. ChemPhysChem 10:1533–40. doi: 10.1002/cphc.200900195

Duncan MJ, Mann EL, Cohen MS, et al (2005) The distinct binding specificities exhibited by enterobacterial type 1 fimbriae are determined by their fimbrial shafts. J Biol Chem 280:37707–16. doi: 10.1074/jbc.M501249200

Forero M, Yakovenko O, Sokurenko E V, et al (2006) Uncoiling mechanics of Escherichia coli type I fimbriae are optimized for catch bonds. PLoS Biol 4:1509–1516. doi: 10.1371/journal.pbio.0040298

Fällman E, Schedin S, Jass J, et al (2004) Optical tweezers based force measurement system for quantitating binding interactions: system design and application for the study of bacterial adhesion. Biosens Bioelectron 19:1429–1437. doi: dx.doi.org/10.1016/j.bios.2003.12.029

Fällman E, Schedin S, Jass J, et al (2005) The unfolding of the P pili quaternary structure by stretching is reversible, not plastic. EMBO Rep 6:52–6. doi: 10.1038/sj.embor.7400310

Hahn E, Wild P, Hermanns U, et al (2002) Exploring the 3D Molecular Architecture of Escherichia coli Type 1 Pili. J Mol Biol 323:845–857. doi: 10.1016/S0022-2836(02)01005-7

Klinth JE, Pinkner JS, Hultgren SJ, et al (2012) Impairment of the biomechanical compliance of P pili: a novel means of inhibiting uropathogenic bacterial infections? Eur Biophys J 41:285–95. doi: 10.1007/s00249-011-0784-2

Lacoursière C (2007) Ghosts and Machines : Regularized Variational Methods for Interactive Simulations of Multibodies with Dry Frictional Contacts.

Lugmaier R a, Schedin S, Kühner F, Benoit M (2008) Dynamic restacking of Escherichia coli P-pili. Eur Biophys J 37:111–20. doi: 10.1007/s00249-007-0183-x

Miller E, Garcia T, Hultgren SJ, Oberhauser AF (2006) The mechanical properties of E. coli type 1 pili measured by atomic force microscopy techniques. Biophys J 91:3848–56. doi: 10.1529/biophysj.106.088989

Mortezaei N, Singh B, Bullitt E, et al (2013) P-fimbriae in the presence of anti-PapA antibodies: new insight of antibodies action against pathogens. Sci Rep 3:3393. doi: 10.1038/srep03393

Mu X-Q, Bullitt E (2006) Structure and assembly of P-pili: a protruding hinge region used for assembly of a bacterial adhesion filament. Proc Natl Acad Sci U S A 103:9861–6. doi: 10.1073/pnas.0509620103

Rangel DE, Marín-Medina N, Castro JE, et al (2013) Observation of Bacterial Type I Pili Extension and Contraction under Fluid Flow. PLoS One 8:e65563. doi: 10.1371/journal.pone.0065563

Sauer F, Mulvey M, Schilling J, et al (2000) Bacterial pili: molecular mechanisms of pathogenesis. Curr Opin … 3:65–72.

Schwan WR, Lee JL, Lenard FA, et al (2002) Osmolarity and pH growth conditions regulate fim gene transcription and type 1 pilus expression in uropathogenic Escherichia coli. Infect Immun 70:1391. doi: 10.1128/IAI.70.3.1391

Servin M, Lacoursière C (2007) Rigid body cable for virtual environments. IEEE Trans Vis Comput Graph 14:783–96. doi: 10.1109/TVCG.2007.70629

Verger D, Bullitt E, Hultgren SJ, Waksman G (2007) Crystal structure of the P pilus rod subunit PapA. PLoS Pathog 3:e73. doi: 10.1371/journal.ppat.0030073

Zakrisson J, Wiklund K, Axner O, Andersson M (2012) Helix-like biopolymers can act as dampers of force for bacteria in flows. Eur Biophys J 41:551–60. doi: 10.1007/s00249-012-0814-8

Zakrisson J, Wiklund K, Axner O, Andersson M (2013) The shaft of the type 1 fimbriae regulates an externalforce to match the FimH catch bond. Biophys J 104:2137–2148. doi: 10.1016/j.bpj.2013.03.059

## FIGURE LEGENDS

**Figure 1.** A) A low resolution 3D reconstruction map of P pili composed of PapA subunits. Subunits are connected head-to-tail via a hinge and interactions between the $n^{th}$ and $(n + 3)^{th}$ subunits stabilize the helix-like structure. For representative purpose we have used the same color for three adjacent subunits B) The rigid multibody model of a pilus. Subunits are held together via head-to-tail bonds whereas layers interact via layer-to-layer interactions (red dots), which are modeled by introducing charges with a given interaction potential. Since the model takes into account the 3D geometry of the structure subunit n will interact with n + 3 (blue dots) when unwound.

**Figure 2.** A) A schematic of an optical tweezers force extension experiment of a pilus. The bacterium and trapped bead is separated adding strain on the pilus. B) A typical force-extension response of a pilus assessed with force measuring optical tweezers. The black curve corresponds the unwinding whereas the red to rewinding. The two curves are of similar shape, except for the dip at around 3.5 - 4 µm that is a consequence of the rewinding an open coiled structure into a helical in which the subunits n and n + 3 interact. In the open coiled structure any such pairs of subunits are too far apart for their charges to interact. The structure needs slack in order for the charges to interact, which thus, in this case, takes place for an elongation of 3.5 µm. The small hysteresis between the curves at the constant force level, region II, indicates that energy is dissipated to the surrounding.

**Figure 3**. Rotational energy potentials plotted using Eq. 4 and values in table 1. The red curve represents the potential in $\theta$-direction and the black curve represent the potential in the $\varphi$- and $\psi$-direction (they are identical) respectively. The inset shows the plot at a different scale.

**Figure 4**. A) Force-extension data from a simulation of the rigid multibody model with 250 subunits. The indices in the figure represent different configurations of the pilus under strain, which are detailed explained in table 2. B) Snapshots of a rigid multibody force-extension simulation. a) The simulation starts without any tension and the pilus resides in a helix-like configuration. d) Extension results in sequential unwinding, starting from left, and the force remains constant throughout this region of extension. e) All *ll*-bonds are broken, only the head-to-tail bonds resist the tension (Region III). f) The pili is fully elongated. j) Tension is reduced and the pilus regains its helix-like structure via sequential rewinding.

**Figure 5.** A) A force-extension data curve from an optical tweezers experiment of a pilus in the presence of antibodies. Several force peaks can be seen that originates from the attachment of antibodies to the shaft subunits, thus claming layers. B) A simulation curve of the rigid multibody model where layer interactions has been changed to show similar respons as experimental data. 30 % of the *ll*-bonds were randomly selected to have a 2 fold increase and a 5 fold decrease during during unwinding and rewinding, respectively. C) A snapshot of the rigid multibody model of 30 subunits where the simulated antibodies are bound to the pili during the unwinding of the structure. Two segments are in this simulation locked by antibodies. The antibodies (blue Y-shaped objects) in this figure were add after the simulation for representative purpose only.

## Figures

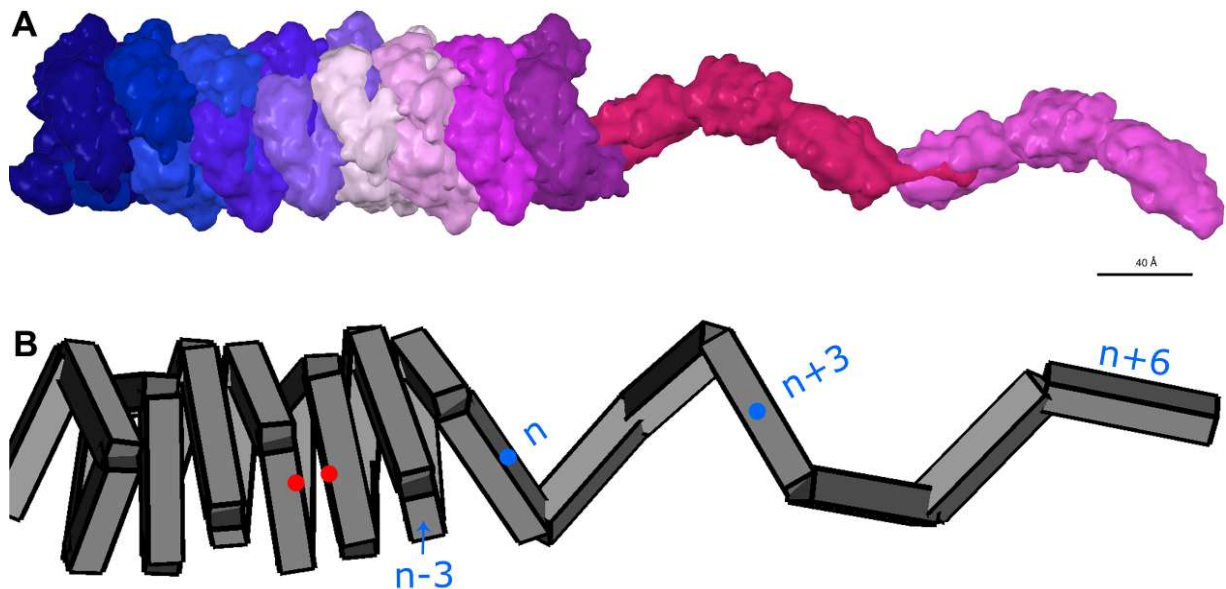

Figure 1

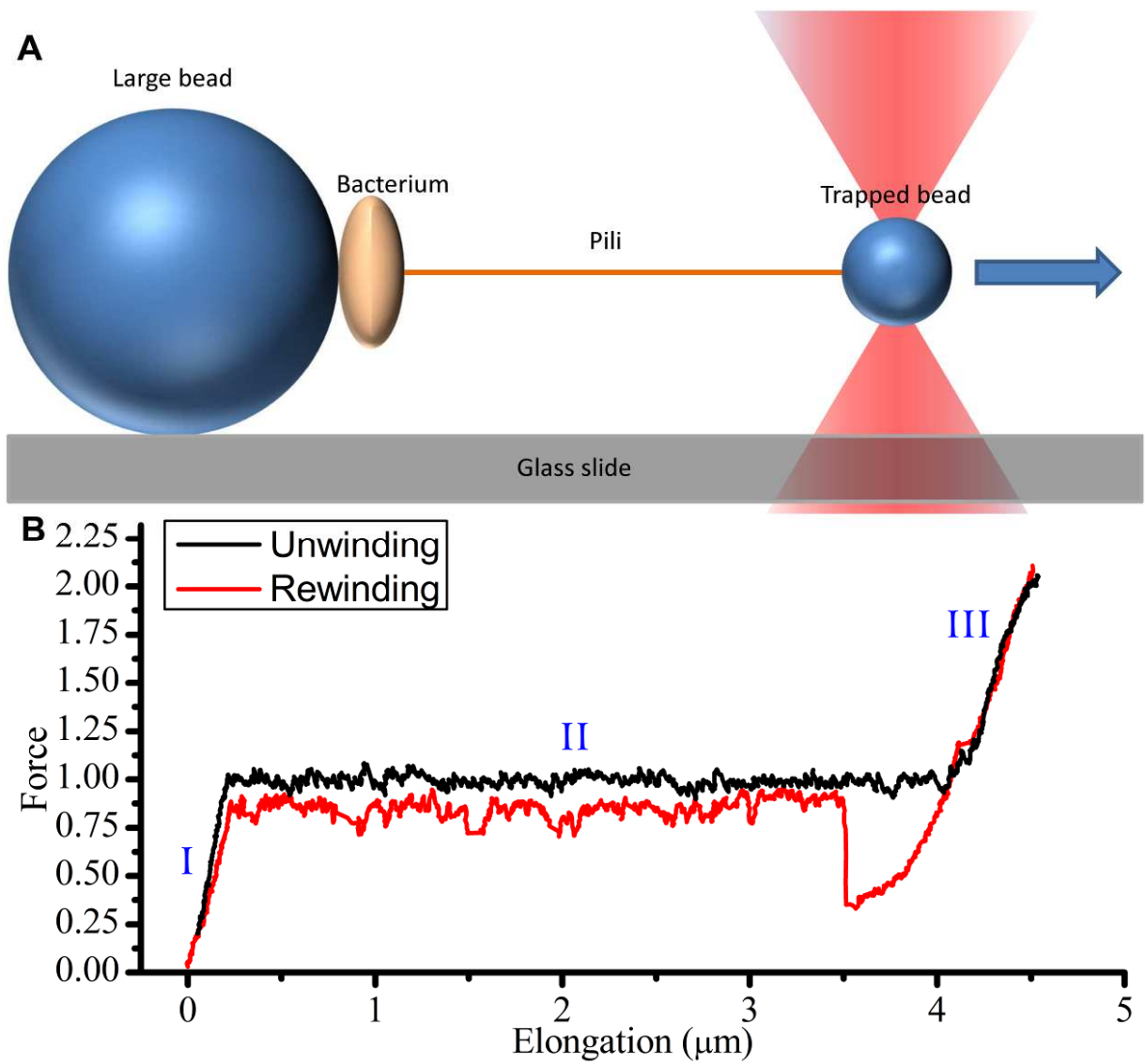

**Figure 2**

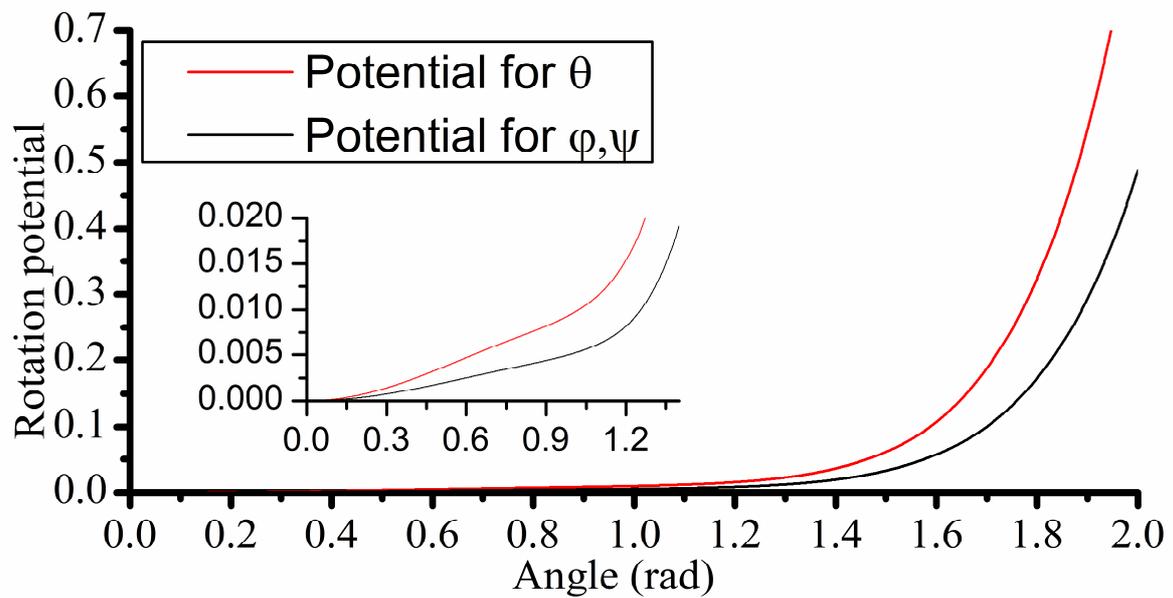

**Figure 3**

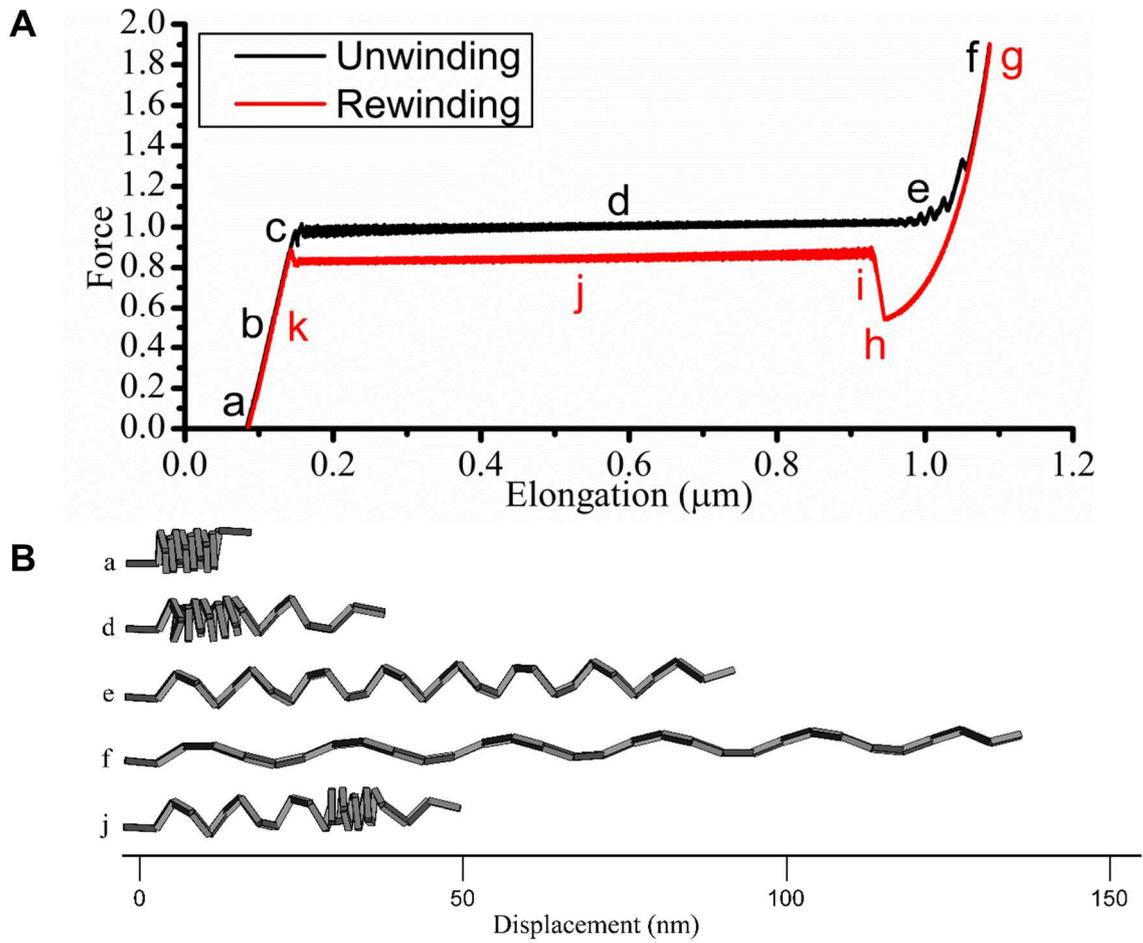

**Figure 4**

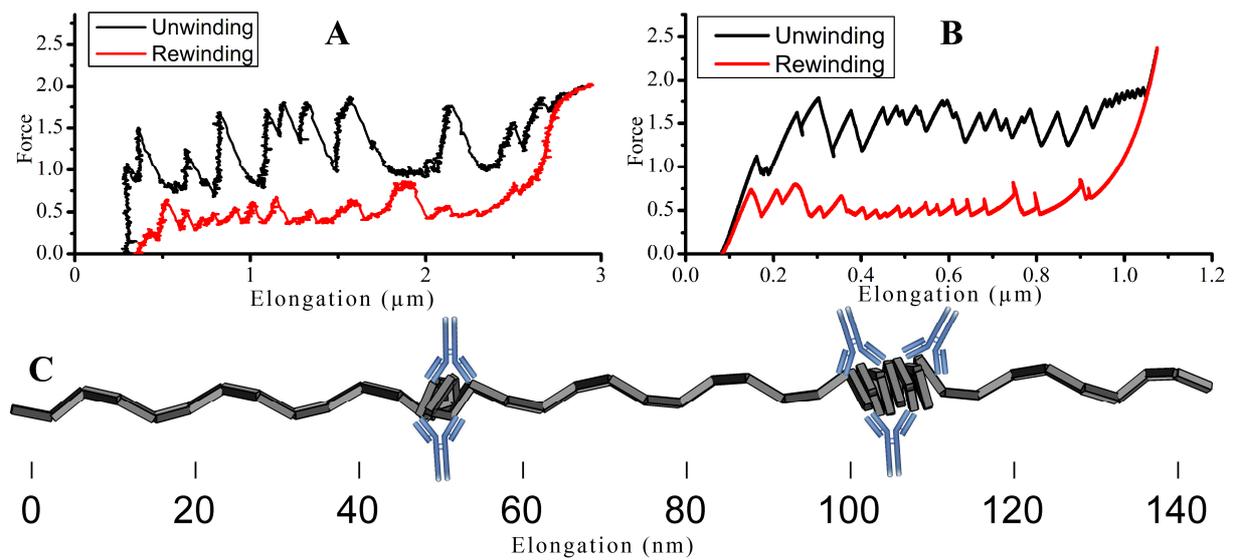

**Figure 5**

## Tables

**Table 1. Coefficients used in Eq. 4 for the case with a Type 1 pili.**

| 2n | $\kappa_{\theta,2n}$ | $\kappa_{\varphi,2n}$, $\kappa_{\psi,2n}$ |
|----|------|------|
| 2 | $1.68 * 10^{-2}$ | $8.94 * 10^{-3}$ |
| 4 | $-1.21 * 10^{-2}$ | $-6.44 * 10^{-3}$ |
| 6 | $5.18 * 10^{-3}$ | $2.77 * 10^{-3}$ |
| 8 | $-1.27 * 10^{-3}$ | $-6.79 * 10^{-4}$ |
| 10 | $1.01 * 10^{-3}$ | $5.39 * 10^{-4}$ |

**Table 2. Description of the different phases seen in Fig. 3B.**

| | |
|---|---|
| **Extension phase** | |
| **a** | The simulation starts with the pilus in a helix-like structure without any tension applied. The laser trap point is moved at a constant velocity, thus leading to a stretching force. |
| **b** | Stretching results in an evenly distributed layer-to-layer separation. In this region, denoted region I, the applied force depends linearly on the extension. |
| **c-e** | The layer-to-layer bonds are sequentially broken as the pilus is stretched. The force remains constant throughout this region of unwinding, region II, besides minor variations due to oscillations in the system that arises when bonds are broken. |
| **e-f** | All layer-to-layer bonds are broken and unwounded. Only the head-to-tail bonds resist tension. The total stretching force is distributed evenly along the pili and the pull force, which has an entropic origin increases non-linearly with the extension. |
| **Contraction phase** | |
| **g-h** | The pilus gradually shortens until two subunits are close enough to feel attraction by the long range layer-to-layer potential. |
| **h-i** | When a first layer is formed, subsequent layers rapidly interact, which leads to a contraction of the pilus and an increase in the force. This nucleation process proceeds until there is a balance between the tension in the pili tension and the force of the laser trap. |
| **i-k** | The pilus refolds sequentially into new layers as the trap is reversed. The force remains constant throughout this process until all subunits are re-folded. |
| **k-a** | The pili is now in the same state as at point *c*, i.e. in helical structure with a minor layer-to-layer separation, and the force- extension curve follows the same linear relation as in *a-c)*. |

**Table 3. Coefficients used in Eq. 4 for the case with a P pili.**

| 2n | $\kappa_{\theta,2n}$ | $\kappa_{\varphi,2n}$ , $\kappa_{\psi,2n}$ |
|----|----------------------|--------------------------------------------|
| 2  | $2.39 * 10^{-2}$     | $3.13 * 10^{-2}$                           |
| 4  | $-1.07 * 10^{-1}$    | $-1.14 * 10^{-1}$                          |
| 6  | $2.25 * 10^{-1}$     | $2.40 * 10^{-1}$                           |
| 8  | $-2.28 * 10^{-1}$    | $-2.43 * 10^{-1}$                          |
| 10 | $1.04 * 10^{-1}$     | $1.11 * 10^{-1}$                           |